\documentclass[twocolumn,prb]{revtex4}
\usepackage{epsfig}

\begin{document}

\title{How to derive and parameterize effective potentials in
colloid-polymer mixtures }

\author {P.G. Bolhuis$^{ \dagger \ddagger}$ and A.A. Louis$^\dagger$\\
$\dagger$ Department of Chemistry, Lensfield Rd,\\ Cambridge CB2 1EW,
UK\\ $\ddagger$ Department of Chemical Engineering, \\Nieuwe
Achtergracht 166, 1018 WV, Amsterdam,\\ The Netherlands}

\begin{abstract}
\noindent 
Polymer chains in colloid-polymer mixtures can be coarse-grained by
replacing them with single soft particles interacting via effective
polymer-polymer and polymer-colloid pair potentials. Here we describe
in detail how Ornstein-Zernike inversion techniques, originally
developed for atomic and molecular fluids, can be generalized to
complex fluids and used to derive effective potentials from computer
simulations on a microscopic level.  In particular, we consider
polymer solutions for which we derive effective potentials between the
centers of mass, and also between mid-points or end-points from
simulations of self-avoiding walk polymers.  In addition, we derive
effective potentials for polymers near a hard wall or a hard
sphere. We emphasize the importance of including both structural {\em
and} thermodynamic information (through sum-rules) from the underlying
simulations.  In addition we develop a simple numerical scheme to
optimize the parameterization of the density dependent
polymer-polymer, polymer-wall and polymer-sphere potentials for dilute
and semi-dilute polymer densities, thus opening up the possibility of
performing large-scale simulations of colloid-polymer mixtures.  The
methods developed here should be applicable to a much wider range
effective potentials in complex fluids.
\end{abstract}

\maketitle {\em this paper has appeared as \\ Macromolecules {\bf 35},
1860 (2002)}

\section{Introduction}
Binary mixtures of colloidal particles and non-adsorbing polymers have
received renewed and growing attention recently, in part because they
exhibit complex and interesting structure, phase behavior, interfacial
properties, and
rheology~\cite{Russ89,Lekk92,Meij94,Ilet95,Verm98,deHo99,Poon01,Lowe01,Yodh01,Loui01a,Evan01},
and in part because they are excellent model systems for the study of
large length and time-scale separations in complex fluids.  Problems
with bridging length-scales are immediately apparent in even the
simplest models of colloid-polymer mixtures: while the mesoscopic
colloidal particles can be modeled as hard convex bodies, the polymers
are often treated at the microscopic (Kuhn) segment level.  Thus, even
though the average size of the polymer coils may be of the same order
of magnitude as that of the colloids, the number of degrees of freedom
needed to model the former may be several orders of magnitude larger
than what is needed for the latter.  This naturally provokes the
question: Can the polymers also be modeled as single composite
particles?  In fact, this is exactly what was done by Asakura and
Oosawa (AO) who, in their classic work on colloid-polymer
mixtures\cite{Asak54,Asak58}, modeled the polymers as ideal particles
with respect to each other, and as hard-spheres with respect to the
colloids.  This model is strictly speaking only valid for
non-interacting polymers, or for interacting polymers in the dilute
limit, while many interesting phenomena such as polymer induced phase
separation take place at finite concentrations of interacting
polymers.  Our ultimate goal, therefore, is to go well beyond the AO
model and describe non-ideal polymers in a good solvent up to
semi-dilute densities.  We recently extended the AO concept to take
into account polymer-polymer interactions, first by rather naively
adding a Gaussian repulsion between the polymers\cite{Loui99a}, and
then by carrying out a much more sophisticated programme which
resulted in {\em density dependent} polymer-polymer and polymer-wall
pair interactions\cite{Loui00,Bolh01}, or, in a complementary
approach, {\em density independent} many-body
interactions\cite{Bolh01a}.

 In this paper we revisit the route to the density dependent pair
potentials in much more detail than in our previous
work\cite{Loui00,Bolh01}.  These density dependent interactions are
derived by inverting structural information, namely the center of mass
(CM) radial distribution function $g(r)$ for the polymer-polymer
interactions, and the wall-polymer or sphere-polymer CM density
profile $\rho(r)$ for the polymer-wall or polymer-sphere interactions
respectively.  We focus here on the considerable subtleties inherent
in these inversions, in particular the importance of using
thermodynamic information or related sum-rules to achieve accurate
potentials.  Many of the lessons learned should be valid for a wider
set of effective potentials.  As a first example, we derive effective
potentials for polymer solutions based on mid-point and end-point
representations.

We also revisit the problem of deriving effective potentials for a
wall-polymer interaction, again emphasizing the subtleties involved in
the inversion process. In addition, we show how to use a similar
method to derive accurate sphere-polymer potentials, a key step
towards large-scale simulations of colloid-polymer mixtures.  Whereas
the integral over the potentials holds the key to accurate
thermodynamics in the homogeneous polymer case, here the relative
adsorption, defined as the integral over the density profile near a
wall or sphere, is the key to achieving the correct thermodynamics.

We also derive a parameterization scheme for the potentials used in
the homogeneous and the inhomogeneous systems.  To achieve this, we
use a novel Monte-Carlo scheme which should be easily adaptable to a
wider class of effective potentials.

 The paper is organized as follows: In Sec.~\ref{sec:model} we repeat
the most important aspects of the models and simulation methods we
used. In Sec.~\ref{sec:pot} we describe how to derive polymer-polymer,
polymer-wall, and polymer-sphere effective potentials from
self-avoiding walk (SAW) polymer simulations, and discuss the
subtleties involved in the inversion procedures.  We also derive other
representations of the polymer-polymer potential, such as the
mid-point or end-point representation.  Section~\ref{sec:fit} contains
the fitting procedures we use to parameterize the density dependent
potentials. We end with the customary concluding remarks.

\section{Simulation Models and Methods}
\label{sec:model}
Simple lattice models, such as the self avoiding walk (SAW), are
widely used to describe flexible, uncharged polymers in a good
solvent.  Their popularity stems in part from the fact that, even
though they ignore all chemical details except excluded volume and
polymer connectivity, these models can still reproduce the scaling
behavior and many other physical properties of athermal polymer
solutions~\cite{deGe79,Shaf00}. In addition, because of their
simplicity, these models are ideally suited for highly efficient
numerical simulation algorithms.
 
The SAW model consists of a cubic lattice of size M on which N chains
of L monomers are placed. Repulsion between (monomer) segments is
built in by allowing only one segment per lattice site.  The monomer
concentration is given by $c=NL/M$, while the polymer chain
concentration is given by $\rho=N/M$.  The size of the polymers is
determined from the radius of gyration $R_g$, which, for an isolated
chain, scales as $R_g \sim L^{0.59}$.

Solutions of flexible polymers in a good solvent can be classified as
a function of polymer concentration into the dilute, semi-dilute and
melt regimes, each with different scaling behavior.  The density or
concentration at which, roughly speaking, the dilute regime crosses
over to the semi-dilute regime is called the overlap concentration or
density $\rho^*$, defined as $\frac{4}{3} \pi \rho^* R_g^3 = 1$, and
corresponding to 1 polymer per unit volume $\frac{4}{3} \pi R_g^3$.
In both these regimes, the monomer concentration $c \ll 1$, and the
only relevant length-scale is $R_g$.  When, upon increasing the
polymer density, the monomer concentration becomes appreciable, the
system crosses over to the so-called melt regime where the monomer
size becomes an additional relevant length-scale.  Thus, when modeling
the semi-dilute regime, it is important to take polymer chains that
are long enough, to ensure that $c$ is still small. The monomer
density $c^*$ at the overlap concentration $\rho^*$ scales roughly as
$c^*\approx 4 L^{-0.8}$ for SAW lattice chains (see
Ref.~\cite{Bolh01}), which implies that for polymers of finite length,
there is only a limited semi-dilute regime.  We studied polymers with
lengths $L=500$ and $L=2000$.  For an isolated $L=500$ coil the radius
of gyration is $R_g =16.5 \pm 0.02$ so that $c=0.027$ at
$\rho=\rho^*$, and $c=0.23$ at $\rho/\rho^*=8.70$, the highest density
we study for this length.  For $L=2000$ chains, $R_g=37.45 \pm 0.04$,
so that $c=0.009$ at $\rho=\rho^*$, and $c=0.064$ at $\rho/\rho^*
=7.04$, the highest density studied for $L=2000$ polymers. This
suggests that while we do not expect the $L=2000$ polymers to exhibit
any significant corrections to scaling behavior for the densities
studied, the $L=500$ polymers may begin to deviate slightly from the
true semi-dilute regime at the highest densities.  We established
previously that the second virial coefficient for two $L=500$ polymers
is close to the scaling limit~\cite{Bolh01}. This indicates that for
low densities, the effective potentials, obtained by the coarse
graining procedure, are very near the scaling limit where properties
only depend on $R_g$ and not on $L$.

The SAW simulations were performed by employing the Monte Carlo pivot
algorithm~\cite{Daut94,pivot} together with translational moves.  This
simple algorithm is sufficient to sample the configuration space at
low concentration. In the semi-dilute regime, i.e.\ for concentrations
$\rho>\rho^*$, we use Configurational Bias Monte Carlo
algorithms~\cite{frenkelbook,dijkstra}.

\section{Deriving effective pair potentials }
\label{sec:pot}

The central theme of the coarse graining procedure advocated here and
in previous work~\cite{Loui00,Bolh01}, is to replace each polymer by a
single particle, interacting with the other polymer particles via an
effective (pair) potential.  For complex particles like polymers,
there is some freedom in the choice of coordinate for the single
``particle''.  We have mainly used the CM, but one could also use the
average monomer, the endpoints, or the midpoint as a monomer based
representation (see e.g.\ the appendix of Ref \cite{Bolh01} or
\cite{Schm01}).  We therefore discuss the midpoint and endpoint
representations in section \ref{sec:midpoints}; a more complete
analysis is planned for a future publication\cite{Loui01}.

\subsection{CM-CM potentials for homogeneous polymer solutions}

In the $\rho \rightarrow 0$, or low density limit, the effective
potentials can be derived from the logarithm of probability of overlap
of the CM of two polymer coils\cite{Loui01a,Bolh01a}.  Calculating
this overlap probability involves integrating over the polymer degrees
of freedom by the Monte Carlo procedures described in
Sec.~\ref{sec:model}.  We sample the configurations of two polymers an
infinite distance apart with the pivot algorithm, and after every 1000
pivot moves we accumulate an overlap probability histogram by testing
for segment overlap as a function of the CM distance.  The effective
potential determined in this manner is approximately Gaussian in
shape, with a range of the order $R_g$ for all lengths.  When the CM's
completely overlap ,the potential has a maximum of $v(r=0)=1.88 \pm
0.01$ for the $L=500$ polymers and $v(r=0)=1.82 \pm 0.02$ for the
$L=2000$ polymers, which is very close to the scaling limit estimate
$v(r=0)=1.80 \pm 0.05$\cite{Bolh01} (Since all interactions are of
entropic origin, we set $\beta = 1/k_BT=1$ throughout this paper). In
the scaling limit the potentials depend only on $R_g$, so that the
probability for complete overlap of the CM's of two polymers will be
independent of their length $L$.

At finite densities there is no longer a simple logarithmic
relationship between overlap probabilities and the effective
potentials\cite{Loui01a,Bolh01a}.  Instead, we use the one--to--one
mapping\cite{Hend74,Reat86} between $g(r)$ and $v(r)$ to find, for
each density, the unique effective potential that exactly reproduces
the two-body correlations.  To generate the pair-correlations, we
performed MC simulations of $L=500$ polymers on a $240 \times 240
\times 240$ cubic lattice, for several different number of polymers
ranging from $N=50$ ($\rho/\rho^* = 0.07$) to $N= 6400$ ($\rho/\rho^*
= 8.70$).  For $L=2000$ polymers we used a lattice of $500 \times 500
\times 500$ units, and the number of polymers ranged from $N=200$
$(\rho/\rho^* = 0.35)$ to $N=4000$ $(\rho/\rho^* = 7.04)$.  During the
simulations we collected the radial distribution functions between the
CM of the polymers. We typically needed on the order of $10^7$
Monte-Carlo moves to achieve sufficient accuracy.

\begin{figure}\epsfig{figure=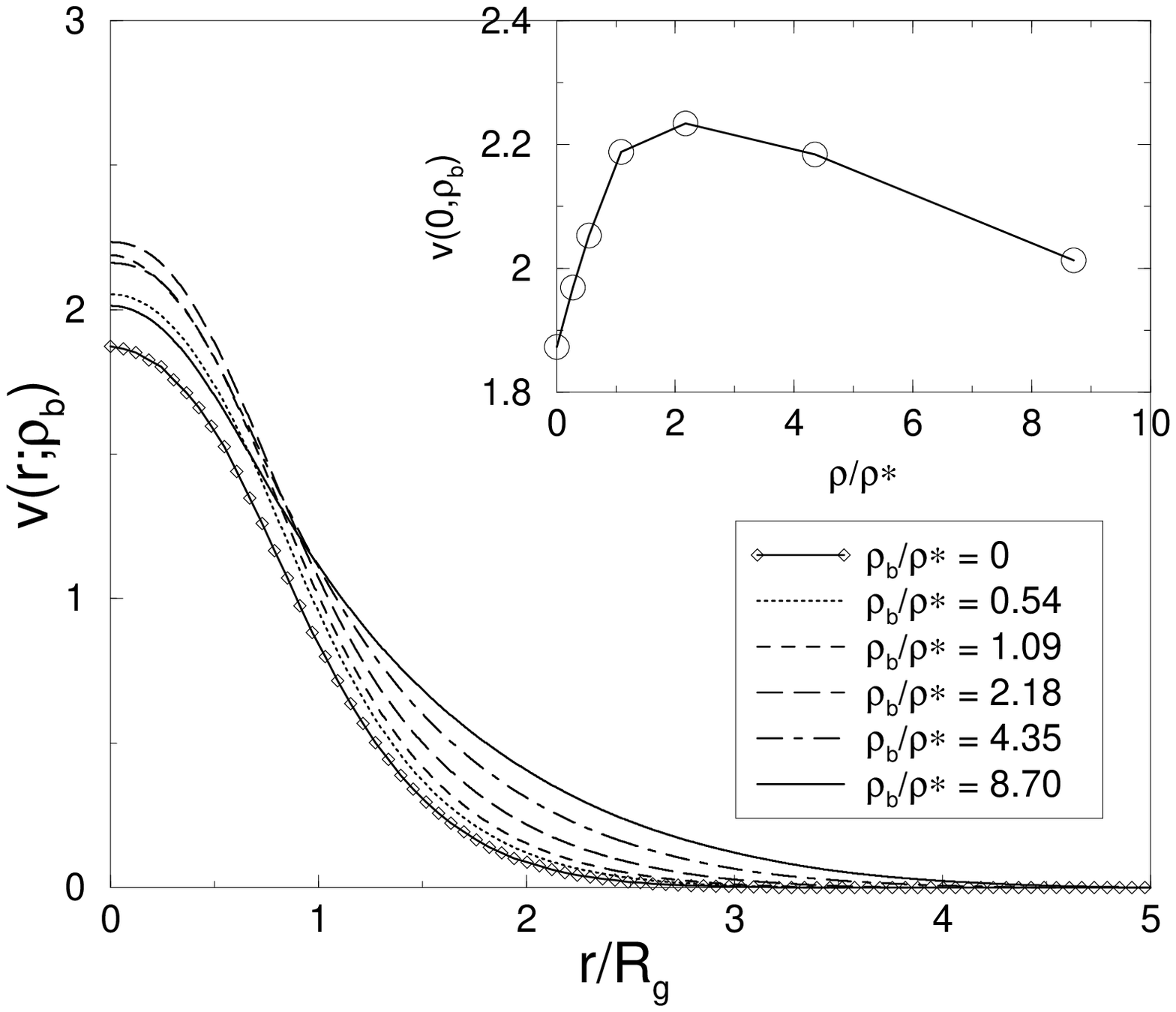,width=8cm} 
\caption{\label{fig:veffL500} The effective polymer CM pair potential
$ v(r;\rho)$ for $L=500$ derived from an OZ-HNC inversion of $g(r)$
for different densities.  The x-axis denotes $r/R_g$, where $R_g$ is
the radius of gyration of an isolated SAW polymer. Inset: The value of
the effective polymer CM pair potential at $r =0$, as a function of
density $\rho/\rho^*$. The maximum of the potential initially
increases before decreasing at high concentration. For clarity we left
out the lowest densities.}
\end{figure}

\begin{figure}\epsfig{figure=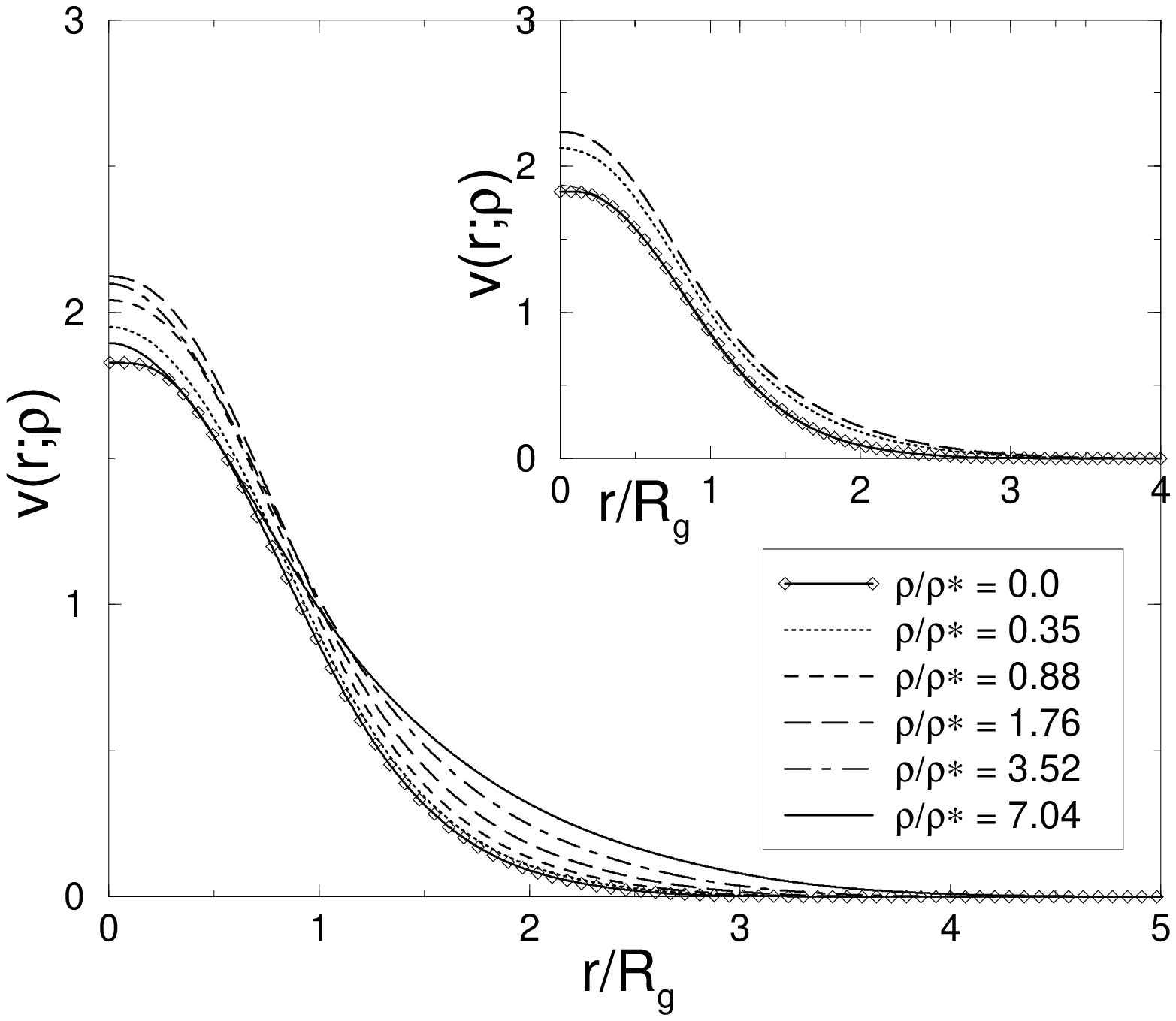,width=8cm} \caption{\label{fig:veffL2000} The effective polymer CM pair potential $
v(r;\rho)$ for $L=2000$.  Inset: A comparison with Figure
~\protect\ref{fig:veffL500} for $\rho/\rho^*=0$ (solid lines), shows
that we are very near the scaling limit.  This is similar for finite
density; e.g. the effective potential for $\rho/\rho^*=1.76$, $L=2000$
(dotted line) is very close to the $\rho/\rho^*=2.18$, $L=500$
potential (dashed line).
}\end{figure}

 The effective potentials $v(r;\rho)$ were constructed using the
one-component OZ equation, supplemented by the hypernetted chain
closure\cite{Hans86}, which is nearly exact for the type of potentials
we generate\cite{Loui00a}.  The resulting effective pair interactions
are plotted in Figure ~\ref{fig:veffL500} and Figure
~\ref{fig:veffL2000} for the $L=500$ and $L=2000$ polymers
respectively, and show a clear though surprisingly small density
dependence.  This density dependence can be understood from the effect
of the density-independent many-body interactions\cite{Bolh01a} which,
for $\rho/\rho^* \leq 1$, are dominated by the three-body
interactions.

 The basic principles of the inversion procedure were already
described in Ref~\cite{Bolh01}, but here we want to point out some
important subtleties that must be kept in mind when performing such
inversions.

\begin{figure}\epsfig{figure=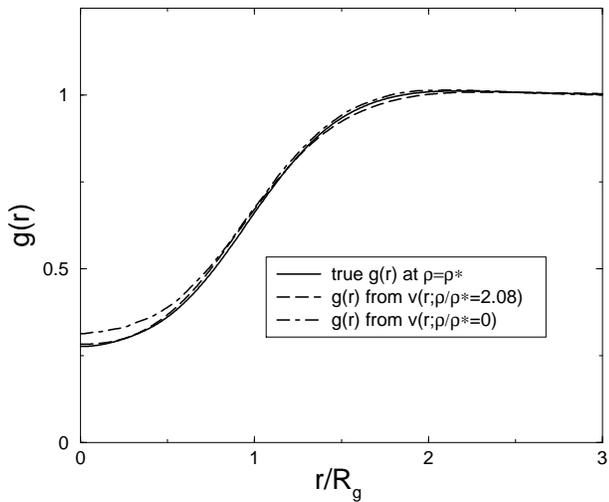,width=8cm} \caption{\label{fig:gofr-rho0} Comparison of $g(r)$'s generated at
density $\rho = \rho^*$ by the low-density potential $v(r;\rho=0)$ and
the higher density potential $v(r;\rho/\rho^* = 2.18)$, compared to
the true $g(r)$ at that density.  Note how small the differences are.
}\end{figure}

 {\bf (1) } First of all, it is important to generate accurate
$g(r)$'s.  This is illustrated in Figure ~\ref{fig:gofr-rho0}, where
two different potentials from Figure ~\ref{fig:veffL500}, namely the
potential at $\rho=0$ and the potential at $\rho/\rho^* = 1.09$, were
used at the same density $\rho/\rho^*=1.09$ to generate the radial
distribution function $g(r)$.  These $g(r)$'s are then compared to the
true SAW $g(r)$ at that density.  Even though the potentials seem
quite different (see Figure ~\ref{fig:veffL500}), the differences in
the radial distribution functions are small, implying that the process
which generates the $g(r)$ must be significantly more accurate than
the difference between the radial distribution functions in Figure
~\ref{fig:gofr-rho0} if one is to obtain accurate effective
potentials.

{\bf (2)} Secondly, because polymers do not have a hard-core, one can
achieve much higher number densities than is normally found for simple
fluids\cite{John64,Reat86}.  This also puts extra demands on the
accuracy of the correlation functions, since at high densities very
small errors tend to destabilize the OZ inversion procedure, making
convergence difficult to achieve.

{\bf (3)} Thirdly, $g(r$) is only known up to half the simulation box
size, rendering the inversion problem underdetermined.  If the
potential is known for all $r$, OZ techniques can be used to extend
$g(r)$ beyond the box-size\cite{Alle87}, but when one is trying to
find $g(r)$ from simulations with an unknown $v(r)$, some assumption
for the large $r$ behaviour of $v(r)$ must be made.  We assume that
$v(r,\rho) =0$ beyond half the box size, but this is not necessarily
obvious at high density, and one needs large simulation boxes to
achieve proper convergence of the effective potentials.  We used boxes
with sides of approximately $14 R_g$.  Although it may seem that the
potentials have vanished already at shorter distances, this is still
near the minimum length necessary to ensure that there are no cut-off
effects in the potentials, especially for higher densities.  Note that
our new Monte-Carlo fitting procedure, described in section
\ref{sec:fit}, partially helps overcome this problem.

\begin{figure}\epsfig{figure=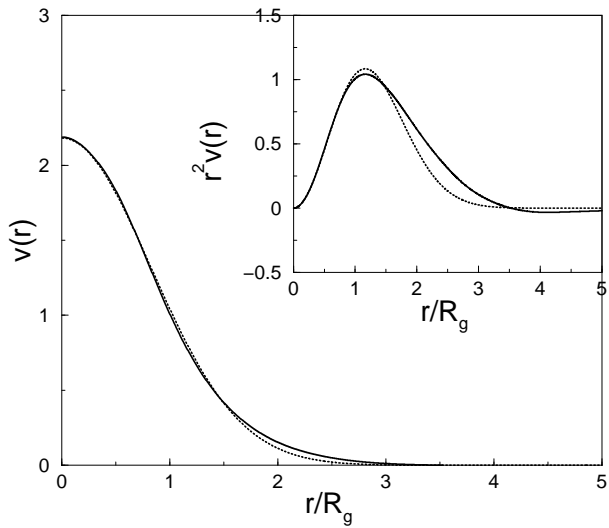,width=8cm} \caption{\label{fig:vrsub} Comparison between the effective potential
$v(r;\rho=\rho^*)$ (solid line) and its fit to a single Gaussian
(dotted line). Although the fit may appear to be quite good to the
eye, the pressures obtained from these potentials differ by about
10\%. The reason for this is illustrated in the inset where the
potential is multiplied by $r^2$, and the differences become more
visible.
}\end{figure}

An effective potential that correctly reproduces pair-correlations
should, in principle, also predict the correct thermodynamics through
the compressibility equation\cite{Hans86}.  But we stress that
correctly determining the equation of state (EOS) $Z =\Pi/\rho$ (here
$\Pi$ is the polymer osmotic pressure) or other thermodynamic
properties from effective potentials can also be quite subtle.  We
illustrate this in Figure ~\ref{fig:vrsub} where we compare a typical
potential and its fit to a single Gaussian.  Although these potentials
do not seem that different, and generate almost identical radial
distribution functions $g(r)$, they result in pressures which differ
by about $10 \%$.  The origin of this difference is clear from the
inset, where we plot $r^2 v(r;\rho)$, which is in fact a better
measure of the relevant differences between the potentials than
$v(r;\rho)$ itself.  That $r^2 v(r;\rho)$ is a better indicator for
the accuracy of the predicted pressure follows from the fact that
potentials such as those shown in Figure ~\ref{fig:veffL500}, Figure
~\ref{fig:veffL2000}, or Figure ~\ref{fig:vrsub} lead to mean-field
fluids\cite{Loui00a,Liko01b}, so named because their equation of state
approximately follows the simple mean-field form:
\begin{equation} 
\label{eq:eos}
Z \equiv P/\rho \approx 1 + \frac{1}{2} \hat{v}(0;\rho) \rho,
\end{equation}
for a wide range of densities.  Here, $\hat{v}(0;\rho)$ is the $k=0$
component of the Fourier transform of the pair interaction, which for
spherical symmetric functions involves the integral over $r^2
v(r;\rho)$.

Keeping these subtleties in mind, we found very good agreement between
the EOS generated by applying the compressibility equation to the
effective potentials in Figure ~\ref{fig:veffL500} and Figure
~\ref{fig:veffL2000} and the EOS of the underlying SAW polymer
solutions.  The compressibility was calculated with the accurate HNC
equation.  The $L=2000$ results almost exactly fall onto the $L=2000$
EOS computed directly by the method of Dickman\cite{Dick87}.  For
$\rho/\rho* < 5$, the $L=500$ EOS calculated with the effective
potentials is very close to the directly computed one, but at higher
densities some small differences develop.  We attribute these to the
difficulties in achieving high accuracy for the tails of the
potentials when the inversions are performed at these higher
densities.  We expect the $L=500$ EOS to be slightly higher than the
$L=2000$ one at higher $\rho/\rho*$ because the monomer density $c$ is
higher for the shorter polymers, which induces $L$ dependent
corrections to the scaling limit at the higher densities.  Both the
$L=500$ and the $L=2000$ EOS, where $c$ remains very small at the
densities probed, are slightly higher than the EOS derived by
renormalization group (RG) calculations\cite{Ohta82,Shaf00}.  They
approximately follow the the correct $\Pi/\rho \sim \rho^{1.3}$
scaling expected for the semi-dilute regime\cite{deGe79} up to the
highest densities.  In contrast, if one were to use the $v(r;\rho=0)$
potential at all densities, the EOS would be underestimated and would
follow mean-field fluid behavior with the incorrect $\Pi/\rho \sim
\rho$ scaling at large $\rho$ instead.  So, even though Figure
~\ref{fig:gofr-rho0} shows that the $\rho=0$ potential results in
pair-correlations $g_2(r)$ that are similar to the true $g_2(r)$'s,
the effective thermodynamics can differ significantly.  Since the
density dependence of the effective pair potentials was shown to arise
from the many-body interactions\cite{Bolh01a}, this immediately
suggests that the corrections to the simple linear $\rho$ scaling of
the EOS are due to the three and higher body interactions.

\subsection{Other representations of potentials for homogeneous polymer solutions}
\label{sec:midpoints}
As mentioned at the beginning of this section, one could also use
other representations for effective pair potentials between polymers.
For star-polymers, for example, the mid-point is a more natural
coordinate\cite{Liko98}.  The $f=2$ arm limit of a star-polymer would
correspond to a normal linear polymer, but in the mid-point
representation.

\begin{figure}\epsfig{figure=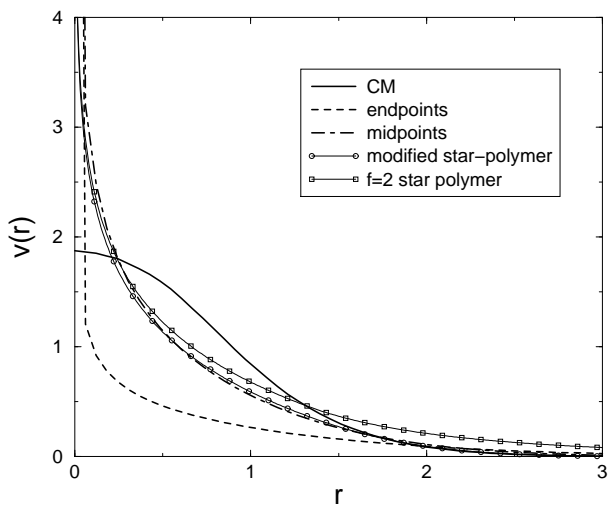,width=8cm} \caption{\label{fig:vr-end} Endpoint, midpoint and CM representations of
the interaction $v(r)$ between two isolated polymers.  Also included
are two fits to the mid-point potentials taken from
Refs\protect\cite{Liko98} and \protect\cite{Dzub00}, as explained in
the text.
}\end{figure}

\begin{figure}\epsfig{figure=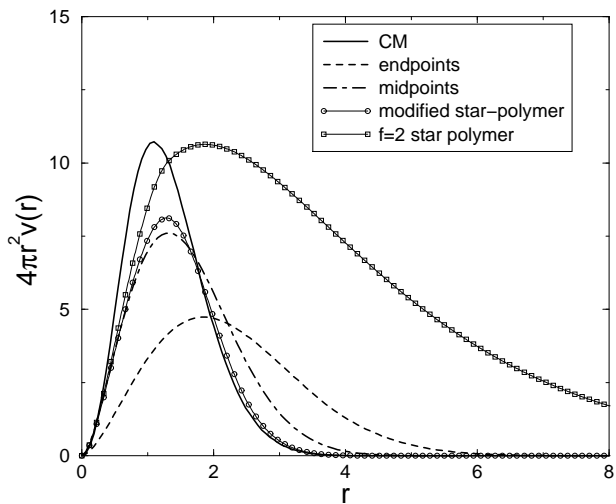,width=8cm} \caption{\label{fig:vr-r2-end} Endpoint, midpoint and CM representations
of the interaction $4 \pi r^2 v(r)$ between two isolated
polymers. Plotting the potentials in this way accentuates the
differences.  Note in particular how poor the $f=2$ limit of the
star-polymer potential\protect\cite{Liko98} performs.
}\end{figure}

In Figure ~\ref{fig:vr-end} we compare the mid-point, end-point, and
CM representations of the interaction between two isolated polymer
coils, all calculated in the usual way by taking the logarithm of the
overlap probability for $L=500$ SAW polymers on a lattice.  We also
compare the $f=2$ arm limit of the star-polymer interaction derived by
Likos {\em et al.}\cite{Liko98}, as well as an improvement especially
tailored for linear polymers\cite{Dzub00}.  The mid-point and
end-point representations have a divergence at full overlap that
scales as $\lim_{r \rightarrow 0} v(r) \propto \ln(r/R_g)$, so that
they appear to be quite different from the CM representation.
However, because the divergence is integrable in 3-dimensions, these
potentials still result in mean-field fluids\cite{Loui01a}.  This is
illustrated in Figure ~\ref{fig:vr-r2-end} where we plot $4 \pi r^2
v(r)$.  The divergence has disappeared, and we see once again that it
is the tails of the potentials, and not the small $r$ behavior, which
matters for the thermodynamics.  While in Figure ~\ref{fig:vr-end} the
differences between the 2-arm limit of the star-polymer potential and
the true mid-point potential are quite small (just fractions of $k_B
T$), Figure ~\ref{fig:vr-r2-end}, together with eq ~\ref{eq:eos},
shows that they will generate dramatically different EOS.  But even
the specially tailored mid-point potential, which appears to almost
perfectly follow the simulations in Figure ~\ref{fig:vr-end}, differs
visibly when plotted as $r^2 v(r)$.  In contrast, for the SAW
simulation, the three different representations, mid-point, end-point,
and CM, should all lead to the same EOS, provided that, as expected,
the volume terms are similar and small\cite{Bolh01,Loui01a}.

These observations demonstrate a more general point, also mentioned in
the previous section: Just because a fit to a potential appears to be
very close to the original potential, does not necessarily mean it
will generate the correct thermodynamic behavior; great care must be
taken to ensure that the fit conserves the right quantities.  (Note
that the fit in Ref.~\cite{Dzub00} was indeed constrained to give the
correct 2nd virial coefficient, which will result in the correct low
density thermodynamics.)

\begin{figure}\epsfig{figure=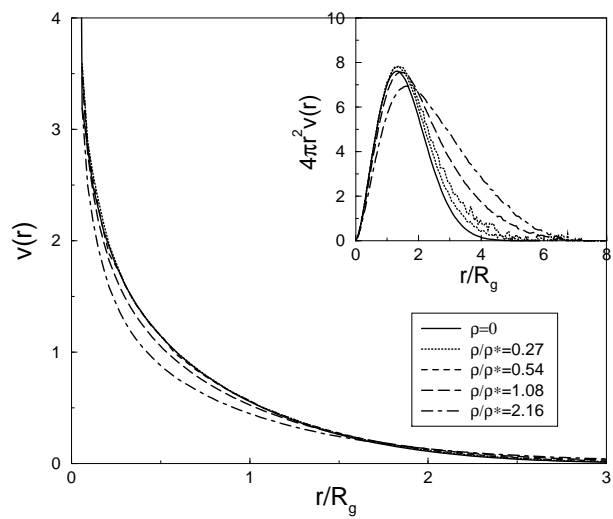,width=8cm} \caption{\label{fig:vr-mid-rho} Midpoint $v(r)$ that exactly reproduces
the midpoint-midpoint $g(r)$ for different bulk polymer densities.
Inset: $4 \pi r^2 v(r)$ shows that the more significant change is in
the tails of the potentials.
}\end{figure}

In principle, one can also derive density dependent potentials for the
mid-point or end-point representations, just as was done for the CM
representation.  In Figure ~\ref{fig:vr-mid-rho}, we show the
mid-point potentials which reproduce the mid-point $g(r)$'s generated
from direct SAW polymer simulations.  We used the same HNC inversion
techniques used earlier for the CM representation, and expect similar
accuracy.  In summary then, while there is some flexibility in the
choice of coordinates, many of the lessons learned for the CM
representation carry right through to the other
representations\cite{Loui01}.

\subsection{Colloid-polymer potentials from wall-OZ and sphere-OZ inversions.}

So far, we have only considered effective polymer-polymer pair
potentials, but a full coarse grained description of a colloid-polymer
mixture, or, more generally, of polymers in confined geometry,
requires effective potentials between polymers and (colloidal)
surfaces as well.  We focus here on non-adsorbing surfaces, and
calculate the effective interaction between polymers and a hard wall,
and between polymers and a hard sphere (HS).  The former is important
for such systems as mixtures of polymers and platelets, or mixtures of
polymers and very large spheres; the latter helps provide a model for
mixtures of spherical colloids and larger polymers.

Near a hard non-adsorbing surface, entropic effects create a polymer
depletion layer, even if the chains themselves are non-interacting.
Although there is no polymer-polymer interaction when modeling such
ideal chains, there will still be a finite wall-polymer potential
$\phi(z)$.  This interaction can be found for ideal polymers near a
wall by the simple result $ \phi(z) = -\ln(\rho(z)/\rho)$, where
$\rho(z)$ is the polymer density at a distance $z$ from the surface
and $\rho$ is the uniform bulk density.  For polymers in an end-point
representation, the interaction reduces to $\phi(z) = - \ln(\mbox{erf}
(z/2R_g))$, while for polymers in the mid-point representation
$\phi(z) = - 2\ln(\mbox{erf}(z/\sqrt{2} R_g))$\cite{Bolh01}.  A
similar result should follow in the CM representation, but we have not
yet succeeded in obtaining an analytic form.

Asakura and Oosawa\cite{Asak54,Asak58} modelled polymers as ideal
w.r.t.\ each other, and with a hard sphere potential $\phi(z)$ of
range $R_g$ w.r.t.\ spheres or walls.  Here we extend their work and
calculate the $\phi(z)$ that exactly reproduces the density profiles
$\rho(z)$ measured by direct simulations of SAW polymers near walls or
spheres.

First we consider SAW polymers near a hard wall.  The effective
 polymer-wall potentials $\phi(z;\rho)$ can be derived from the
 $\rho(z)$ by combining the wall Ornstein Zernike equation with the
 HNC closure, leading to
\begin{equation}
\label{eq:stells}
  \phi(z;\rho) = -\ln(\rho(z)/\rho) + \rho \int d{\bf r'}
\left(\rho(z')/\rho -1 \right) c_b(|{\bf r} - {\bf r'}|).
\end{equation}
Here, $c_b(r)$ is the bulk polymer direct correlation function.  The
details of this inversion procedure are given in
Ref.~\cite{Bolh01}. In contrast to the homogeneous case, the inversion
involves only one iteration, because the bulk polymer-polymer
potentials and hence $c_b(r)$ are already known.

\begin{figure}\epsfig{figure=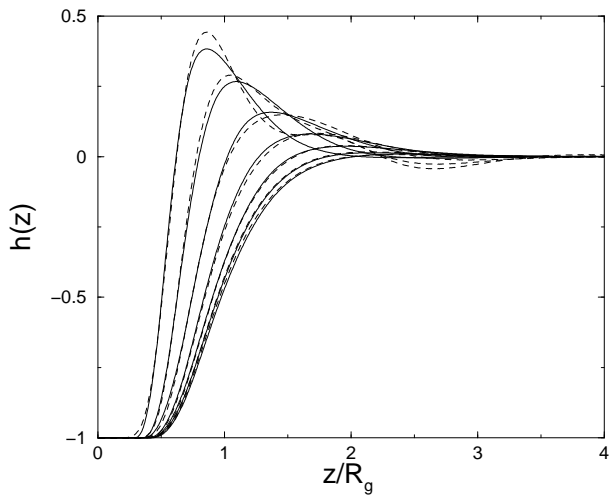,width=8cm} \caption{\label{fig:hz}Measured $h(z)$ (solid lines) next to a wall
obtained from SAW simulation compared with the $h_f(z)$ which follows
from the optimized fitted potential $\phi_f(z;\rho)$ (dashed
lines). From left to right the curves correspond to bulk polymer
densities $\rho/\rho^* = 2.27, 1.16, 0.59, 0.30 ,0.16, 0.08$ and $0$,
respectively. Note that the densities differ slightly from the
corresponding bulk density due to the depletion at the wall.
}\end{figure}

\begin{figure}\epsfig{figure=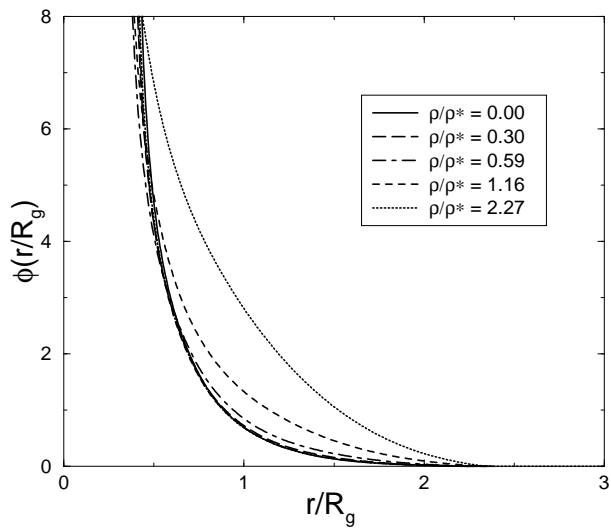,width=8cm} \caption{\label{fig:vz} The wall-polymer potential $\phi(z;\rho)$ as
obtained from the inversion of density profile $\rho(z)$.
}\end{figure}

Using the same explicit SAW polymer model as in
Section~\ref{sec:model}, we performed MC simulations of polymers of
length $L=500$ on a lattice of size $M=240\times 240\times 240$ with
hard planar walls at $x=0$ and $x=240$. The polymer segments were not
allowed to penetrate the walls. The simulations were done for
$N=0,100,200,400,800,1600$, and $3200$. During each simulation, we
computed the density profiles $\rho(z)$, where $z$ denotes the
distance between the polymer CM and the wall.  The normalized profiles
$h(z)=\rho(z)/\rho -1$, for different bulk concentrations $\rho$, are
shown in Figure ~\ref{fig:hz}.  The wall-polymer potentials
$\phi(z;\rho)$ were obtained using the wall-OZ-HNC inversion procedure
and are plotted in Figure ~\ref{fig:vz}.  The wall is essentially
still hard, but is now softened by an additional exponentially
decaying repulsion.  The density dependence is more pronounced than in
case of the effective potentials for bulk polymer solutions.

\begin{figure}\epsfig{figure=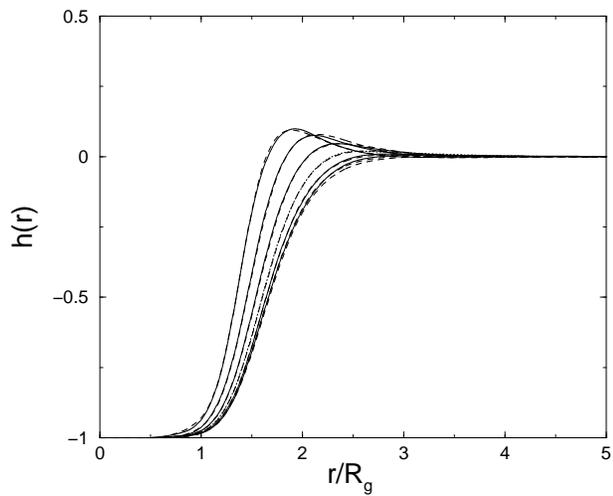,width=8cm} \caption{\label{fig:hrrad16} The polymer density profile $h(r) =
\rho(r)/\rho$ around a colloidal HS of diameter $\sigma = 2 R_g$. From
left to right the curves correspond to a bulk polymer density
$\rho/\rho* =2.18, 1.09, 0.54, 0.27, 0.14$ and $0$, respectively. The
solid lines represent the SAW simulation results, whereas the dashed
lines correspond to the $h(r)$ that results from an optimized fit to
the effective potential $\phi(r;\rho)$.
}\end{figure}

\begin{figure}\epsfig{figure=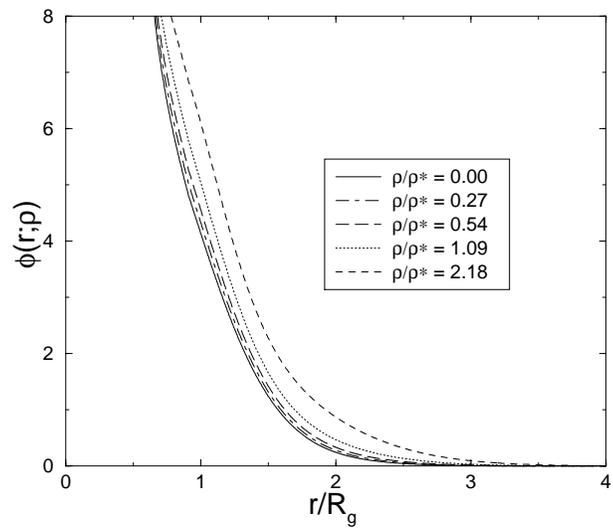,width=8cm} \caption{\label{fig:vrrad16} The colloid-polymer effective potential
$\phi(r)$ for a colloidal diameter of $\sigma 2= R_g$.
}\end{figure}

The potential for a polymer coil interacting with a colloidal HS of
diameter $\sigma$ can be found from the density profile $\rho(r)$
around the sphere. Here $r$ is the distance from the center of the
colloidal HS.  The inversion procedure is very similar to that of the
wall-polymer potential geometry.  For a binary mixture of two
components labeled c and p, in which colloidal component c is
infinitely dilute $(\rho_c \rightarrow 0)$, the binary OZ equations
decouple, and reduce to
\begin{equation}
h_{cp}(r) = c_{cp}(r) +\rho \int d{\bf r'} h_{cp}({\bf r'}) c_b(|{\bf
r} - {\bf r'}|).
\end{equation}
Where $h_{cp}(r) =\rho(r)/\rho -1$, and $c_b(r)$ is again the bulk
polymer direct correlation function which is determined
independently. The above equation can be combined with the HNC closure
to yield
\begin{equation}  \phi(r;\rho) =  -\ln(\rho(r)/\rho) + \rho \int d{\bf r'} h_{cp}( r')
c_b(|{\bf r} - {\bf r'}|).
\end{equation}
which can be solved in one step by Fourier transformation.
Figure~\ref{fig:hrrad16} shows the polymer density profiles
$h_{cp}(r)$ around a single colloidal HS particle with a diameter of
$\sigma=2 R_g$. The effective potentials are plotted in Figure
~\ref{fig:vrrad16}. The interaction appears to be somewhat softer than
for the planar wall case, at least in the sense that it is still
finite at distances that correspond to the CM being ``inside'' the
hard colloidal particle.  This happens more readily for smaller
size-ratios $\sigma/(2 R_g)$ because the polymers can deform around
the colloids.  Note that this penetration into the colloidal HS will
not occur in the mid- or end-point representations.

\subsection{Connection to scaling theory approaches}

Polymers in the semi-dilute regime have traditionally been very
successfully understood with scaling theories\cite{deGe79}.  There the
fundamental unit is no longer the radius of gyration $R_g$ at zero
density, but the blob-size $\xi(\rho)$\cite{deGe79,Shaf00}, which
appears both in the EOS, where $\Pi_b \sim \xi(\rho)^{-3}$, and in the
relative adsorption $\Gamma/\rho \sim \xi(\rho)$.  For the homogeneous
polymer solutions, it is not clear from either the $g(r)$ or from the
effective potentials where the blob-size comes in.  The reason the
effective potentials correctly reproduce the EOS is because they
correctly reproduce the pair-correlations, which, through the
compressibility theorem\cite{Hans86}, give the correct thermodynamics.
This blob-size only appears in the potentials if one attempts to
parameterize them for $\rho/\rho*>2$ (roughly speaking the density at
which the scaling sets in), since the integral over $r^2 v(r)$ should
scale as $\rho^{-2}\xi(\rho)^{-3}$.  Since we only parameterize the
potentials for lower densities, this scaling does not apply to the
functional form we use.

  For polymers near a wall the width of density profile of the
monomers scales with $\xi(\rho)$ in the semi-dilute
regime\cite{deGe79}.  A similar scaling is expected to hold for the
width of the CM density profile.  So even though $\xi(\rho)$ is not
immediately evident in the effective wall-polymer potentials, it is
present in the induced density profiles.  Nevertheless, many questions
as to the direct relationship between the scaling theories and our
soft-colloid approach remain to be worked out.  Partially for that
reason, some care must be taken when applying the soft-colloid picture
deep in the semi-dilute regime.

The semi-dilute regime scaling theories are not normally valid in the
dilute regime, where we expect our soft-colloid approach to be robust.

\section{Fitting effective pair potentials}
\label{sec:fit}

\subsection{CM-CM  potentials for bulk polymer solutions}
\label{sec:bulkfit}
When applying effective potentials in coarse grained simulations or in
theoretical analysis, it is convenient to have an explicit expression
or parameterization of the potentials.  In Ref.\cite{Bolh01} we fitted
the effective pair potentials we obtained by inversion of the $g(r)$
to a single Gaussian and described the remainder by means of a spline
fit.  However, for practical use in simulation or for a theoretical
analysis, one would like to use an analytic expression without the
complication of a multi node spline fit. Moreover, it would be very
useful to be able to explicitly model the density dependence of the
potentials as well.  Since we are dealing with an approximately
Gaussian form, the simplest analytical expression is a sum of
Gaussians with density dependent coefficients:
\begin{equation}
\label{eq:gauss}
v_f(r;\tilde{\rho}) = \sum_{i=0}^n a_i(\tilde{\rho})
e^{-(x/b_i(\tilde{\rho}))^2},
\end{equation}
where we introduced $\tilde{\rho} \equiv \rho/\rho^*$ for clarity.
Here, the Gaussians are centered at $r=0$, because the maximum of the
potential is located at the origin. The density dependence comes in
through the coefficients $a_i(\tilde{\rho})$ and
$b_i(\tilde{\rho})$. In first instance, we take the coefficients to be
linear in density, i.e.  $a_i(\tilde{\rho}) = a_{i0} +
a_{i1}\tilde{\rho}$ and $b_i(\tilde{\rho}) = b_{i0} +
b_{i1}\tilde{\rho}$.

  Note that an expression of this form cannot describe the slightly
negative tails in the potential that were found in Ref.\cite{Bolh01}.
However, as was argued in the previous section, the structure is not
very sensitive to small changes in $v(r;\tilde{\rho})$. A more
important aspect of the fitting procedure is to make sure that the
fitted potentials lead to the correct EOS. To do this we make use of
the accurate mean field approximation in eq ~\ref{eq:eos}. Because
this is such a good approximation, keeping the $k=0$ Fourier component
$\hat{v}(0;\tilde{\rho}) = 4 \pi \int r^2 v(r;\tilde{\rho})$ for the
fitted potential $v_f(r;\tilde{\rho})$ equal to that of the original
potential $v(r;\tilde{\rho})$ ensures that the pressures will also be
(almost) equal.  A fitting procedure should therefore include the
constraint
\begin{equation}
\label{eq:const}
\hat{v}(0;\tilde{\rho}) = \hat{v}_f(0;\tilde{\rho}) =
 \pi^{\frac{3}{2}} \sum_{i=0}^n a_i(\tilde{\rho})
 \left(b_i(\tilde{\rho})\right)^{3},
\end{equation}
which results in one of the coefficients being fixed by the
constraint.  The $\hat{v}(0;\tilde{\rho})$ itself is a smooth function
of the bulk density $\tilde{\rho}$; we fit it to a cubic polynomial.

The inverted potentials $v(r,\tilde{\rho})$ were fitted to eq
~\ref{eq:gauss} by applying a standard least squares non-linear
fitting procedure~\cite{NumRec} including the above constraint. To
obtain a good fit of all the potentials for $\rho/\rho^*<2$ we needed
at least 3 Gaussian terms, particularly to correctly describe
$v_f(r=0;\tilde{\rho})$.  For higher densities, the linear density
dependence in eq ~\ref{eq:gauss} broke down. This can partly be
understood from the fact that the potential at $r=0$ increases almost
linearly for $\rho/\rho^*<2$ but decreases for higher concentration
(see inset of Figure ~\ref{fig:veffL500}). Clearly, a linear fit in
the density cannot cope with this behavior. Higher order polynomials
for the coefficients were not very successful either. A better option
would be to fit the potentials for high densities independently, and
then ensure continuity at the crossover.

After using the non-linear fitting procedure to directly fit the
potentials, the best fit coefficients in eq ~\ref{eq:gauss} can be
further fine-tuned by minimizing the difference between the radial
distribution function $g(r)$ of the SAW polymers and the $g_f(r)$
generated from the fitted potential $v_f(r)$. This minimization
procedure works as follows.  For a certain density
$\tilde{\rho}=\rho/\rho^*$, we first calculated $g_f(r)$ for
$v_f(r;\tilde{\rho})$ with the best fit coefficients using the HNC
approximation. We then compare this $g_f(r)$ to the original radial
distribution function $g_{SAW}(r)$, measured in the explicit SAW
simulation at the same density, by determining the least square
difference: $\Delta(\tilde{\rho}) = \int (g_f(r) -g_{SAW}(r))^2 r^2 dr
$.  We performed this calculation for every density $\tilde{\rho}_i$
we have SAW data for, summing up to $\Delta_{total} = \sum_i
\Delta(\tilde{\rho}_i)$. Subsequently, we changed one of the
coefficients by a small amount and determined the difference
$\Delta_{total}$ again. If the difference between the fitted $g_f(r)$
and the SAW data $g_{SAW}$ becomes smaller, the new coefficient was
accepted, otherwise the old coefficient was restored.  These steps
were repeated, until the difference reached a minimum value and
changes in the coefficients were no longer accepted.  The final
coefficients only represent a local minimum, so this Monte Carlo type
minimization procedure cannot be applied without a good initial
estimate for the fit coefficients.. On the contrary, it depends
strongly on the quality of the original non-linear fit described
above.  Note that by using this optimization we skip one step, namely
the inversion of the original $g(r)$.  If a small error was made in
the inversion, for example by our assumption that $v(r)=0$ for
$r>r_{cutoff}$, it should be corrected during the optimization.
However, one must keep in mind that the minimization procedure only
finds a local minimum.  Larger errors may not be corrected.

We applied the above fitting procedure to the bulk polymer results for
$L=500$ and $L=2000$ . The coefficients are given in
Table~\ref{table1} and Table~\ref{table2}, respectively.  One of the
coefficients, $b_3(\tilde{\rho})$, is fixed by the constraint of eq
~\ref{eq:const}:
\begin{equation}
\left(b_3 (\tilde{\rho})\right)^3 = {a_3(\tilde{\rho})}^{-1} \left[
\pi^{-\frac{3}{2}} \hat{v}(0;\tilde{\rho}) - \sum_{i=1}^{2}
a_i(\tilde{\rho}) \left(b_i(\tilde{\rho})\right)^3 \right].
\end{equation}
Note that this implies that in contrast to the other coefficients,
 $b_3(\tilde{\rho})$ is no longer linear in density.  The best fit
 coefficients for both lengths are not far from each other, again
 indicating that the results are close to the scaling limit.

\begin{figure}\epsfig{figure=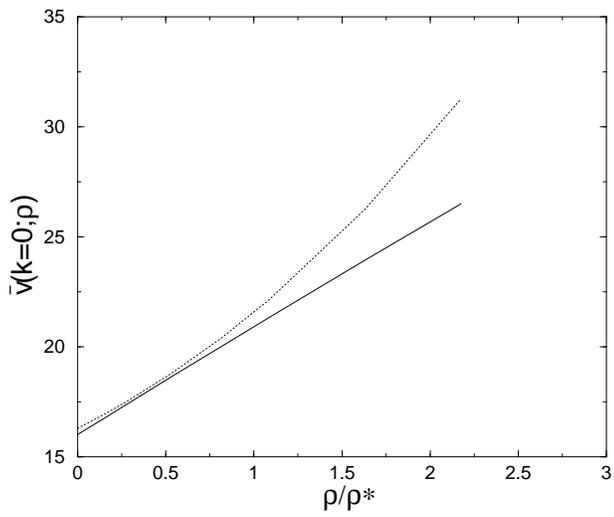,width=8cm} \caption{\label{fig:compvhat} When the explicit constraint of eq
~\protect\ref{eq:const} is not included, the value of
$\hat{v}(k=0;\rho) = 4 \pi \int r^2 v(r;\rho)$ (dotted line) begins to
deviate significantly from the correct value (solid line).  This will
have an important effect on the EOS.
}\end{figure}

To illustrate the importance of explicitly including the constraint of
eq ~\ref{eq:const}, we applied the same non-linear fitting procedure,
including the Monte-Carlo optimization, but without the constraint.
In Figure ~(\ref{fig:compvhat}) we compare $\hat{v}(0;\tilde{\rho})$,
which is a good measure for the EOS, for the constrained and the
unconstrained case. Clearly, including the constraint is essential for
reproducing the correct thermodynamic behavior.  At this point we
would like to return to a statement made in an earlier
paper\cite{Bolh01}, where we emphasized the importance of the very
small negative tails we found from our inversion procedure.  Because
the tails are so small, their effect on the radial distribution
functions is almost imperceptible.  It is only when they are
multiplied by $r^2$ that they become important for the EOS.  Our
current fitting procedure ignores the negative tails, but because we
force the sum-rule on $r^2 v(r)$, the thermodynamics are still
correctly reproduced. Therefore our previous statements on the
importance of the tails should be tempered.  It is actually the
sum-rule on the pair-potential which is critical to achieving the
correct thermodynamics.

\subsection{Colloid-polymer potentials}

A similar fitting procedure as in the previous section can be applied
to the effective wall-polymer potentials.  In Ref.~\cite{Bolh01} we
showed that the effective wall-polymer potential
$\phi(z,\tilde{\rho})$ can be fitted reasonably by the cubic
exponential function
\begin{equation}
\label{eq:hzfit}
\phi_f(z,\tilde{\rho}) = a_0(\tilde{\rho}) \exp \left[
a_1(\tilde{\rho}) z + a_2(\tilde{\rho}) z^2 +a_3(\tilde{\rho}) z^3.
\right]
\end{equation}
As before, the density dependence can be introduced through the
coefficients $a_i(\tilde{\rho}) = a_{i0} + a_{i1}\tilde{\rho} +
a_{i2}\tilde{\rho}^2 $. Here we chose a quadratic density dependence,
because a linear expression yielded a poor fit. We fitted the
potentials $\phi(z;\tilde{\rho})$ using the non linear fit method of
the previous section but without any constraint. Subsequently, the
coefficients were optimized using the minimization procedure mentioned
above.  The optimized best fit coefficients are given in
Table~\ref{table3}, and the $h_f(z)$ generated by these fitted
potentials are shown in Figure ~\ref{fig:hz}.  The fit seems to be
reliable for $\rho<\rho^*$, but, unfortunately, begins to break down
in the semi-dilute regime. For the highest densities, the reproduced
$h_f(z)$ show more structure than the density profile obtained from
the SAW simulations. Clearly, eq ~\ref{eq:hzfit} cannot completely
capture all the wall-polymer effects in the semi-dilute regime.  The
functional form of eq ~\ref{eq:hzfit} is rather {\em ad hoc}, and the
fit would probably be better if one had access to a more accurate
analytical expression based on physical arguments.

\begin{figure}\epsfig{figure=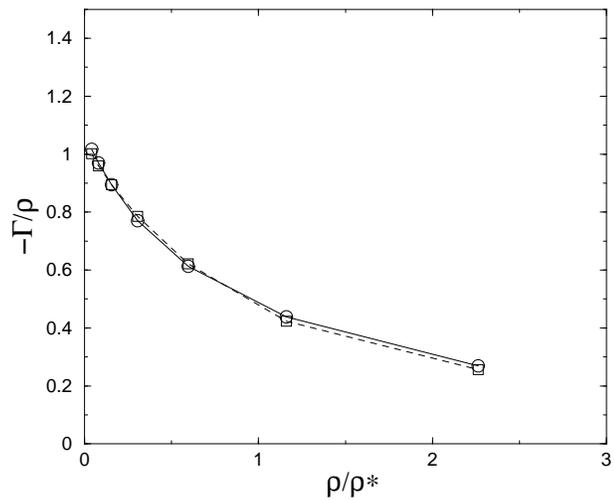,width=8cm} \caption{\label{fig:gamma} The relative adsorption $\Gamma/\rho$ (in
units of $R_g^{-2}$) given by direct SAW simulations of $L=500$
polymers next to wall (open circles), compares well to the adsorption
calculated from the fitted wall-polymer potentials (open squares).
}\end{figure}

Similarly to the bulk polymer case, one needs to compare surface
thermodynamic properties to ensure the quality of the fit.  We focus
on the polymer adsorption at the wall, $\Gamma$, defined as the
partial derivative of the excess grand potential $ \Omega^{ex}$ per
unit area with respect to the chemical potential $\mu$,
\begin{equation}\label{eq:adsorption}
\Gamma = -\frac{\partial( \Omega^{ex}/A)}{\partial \mu} = \rho
\int_0^{\infty} h(z) dz,
\end{equation}
where $A$ is the surface area.  In Figure ~\ref{fig:gamma} we show
that the adsorption $\Gamma$ is indeed well represented by the fits.
So, even though for high densities the structure next to the wall is
not well described by eq ~\ref{eq:hzfit}, the integral over $h(z)$ is
still accurately represented.  Since the adsorption $\Gamma$ is given
by the integral over $h(z)$, instead of an integral weighted by $z^2$,
the tails of the potentials are less important to the correct
thermodynamics than in the case of the bulk polymer potentials.  In
fact, if the adsorption $\Gamma$ and the EOS are known as a function
of the density, then the surface tension of the polymer solution can
be calculated with the Gibbs adsorption equation\cite{Mao97}.
Therefore, since the fitted wall-polymer potentials give the correct
$\Gamma$, and the polymer-polymer potentials give the correct
$\Pi_b/\rho$, when used together they should provide an accurate
representation of surface tensions and related surface thermodynamic
quantities.

\begin{figure}\epsfig{figure=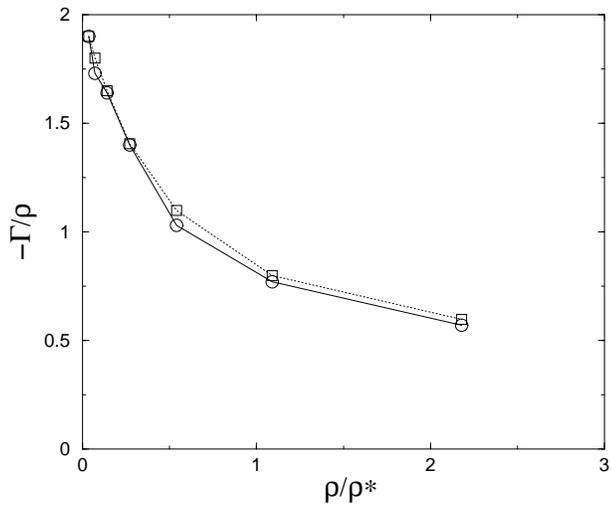,width=8cm} \caption{\label{fig:gamma2} The relative adsorption $\Gamma\rho$ (in
units of $R_g^{-2}$) given by direct SAW simulations of $L=500$
polymers around a colloidal sphere of diameter $\sigma = 2 R_g$ (open
circles), compared to the adsorption from the fitted colloid-polymer
potentials (open squares).  Note that the agreement is not as good as
in the polymer wall case, because of the integration over $ h(r) r^2$
}\end{figure}

In the case of a spherical particle, the fit to the sphere-polymer
potential is more straightforward.  Once again we can use a sum of
Gaussians to estimate the potential
\begin{equation}
\label{eq:colpolgauss}
\phi_{f}(r) = \sum_{i=0}^n a_i(\tilde{\rho})
e^{-((r-c_i(\tilde{\rho}))/b_i(\tilde{\rho}))^2}.
\end{equation}
Here, because of the size difference between the colloidal particle
and the polymer coil, we allow the Gaussians to be off center. In the
non-linear fit for the $\sigma=2 R_g$ data, we set $n=2$ and
$c_1(\tilde{\rho})=0$, and assumed a linear density dependence of the
coefficients. Subsequently, the coefficients were optimized using the
minimization procedure. The results are given in Figure
~\ref{fig:hrrad16} and Table~\ref{tab:rad16}.  The $h_f(r)$ reproduces
the measured $h(r)$ quite well.  To check the quality of the fit we
compared in Figure ~\ref{fig:gamma2} the adsorption of the SAW
polymers around the sphere with the optimized potentials.  Here, in
contrast to the planar wall case, we do have to integrate over $h(r)
r^2$. We would therefore expect that the deviations are larger than
for the wall. However, although the agreement is not as good as in the
wall case, the adsorptions are still well represented, thus giving us
confidence that the fits are adequate.

\section{Conclusion}

In this paper we have shown in detail how to use OZ techniques to
derive effective potentials potentials for polymer solutions from
direct simulations of SAW polymers.  Just as many subtleties were
found in the original application of OZ inversion techniques to simple
atomic and molecular fluids, so we find that great care must be taken
to correctly invert and fit our effective potentials.

We found that $g(r)$ is not very sensitive to differences in $v(r)$,
which is very similar to the situation for atomic and molecular
fluids\cite{Hans86}.  This insensitivity places strong demands on the
accuracy of the original $g(r)$'s needed as input for our inversion
procedures.

For the polymer-polymer potentials, it is crucial that the integral
over $r^2 v(r;\rho)$ is correctly represented.  Seemingly very small
differences in the tails of $v(r)$, which in turn result in almost
imperceptible changes in $g(r)$, can nevertheless result in large
differences in the EOS.  This principle should hold not only for
linear polymers in the CM, end-point or mid-point representations, but
also for low arm-number star polymers\cite{Liko98},
dendrimers\cite{Liko01a}, and other mean-field fluids.

We also derived density-dependent wall-polymer and sphere-polymer
potentials by directly inverting the one-particle density profile
$\rho(r)$ calculated by direct simulations of $L=500$ SAW polymers.
Here, the important thermodynamic parameter is the adsorption
$\Gamma$, which is quite well reproduced by our fitted potentials.
Since our effective potentials provide a good representation of the
EOS and of the adsorption, they should lead to an accurate
representation of the surface tension and other related surface
thermodynamic properties of polymer-colloid mixtures.

And finally, while one might think that most of the hard work is done
once effective potentials have been inverted from direct simulations,
fitting these potentials for their use in large-scale simulations of
colloid-polymer mixtures is far from trivial.  We showed how to use a
numerical optimization procedure to ensure the accuracy of our fits.
Because this optimization procedure skips the direct inversion step,
it can remove residual errors in the inversions, guarantee that the
fits conserve the right sum-rules, and lead to the correct
thermodynamics.  We emphasize that these conclusions should hold for a
much broader class of effective potentials than the ones we discussed
here.

The ultimate goal of our research project\cite{Loui00,Bolh01,Bolh01a}
is to model large-scale mixtures of polymers and colloids.  The
coarse-graining of the polymers from the ``microscopic'' SAW chains to
single composite entities interacting via effective density dependent
pair potentials is a crucial step toward that goal.  The next step
will be to use the accurate fits derived in this paper to perform
direct simulations of many spherical colloids interacting with many
polymers.

\section*{Acknowledgments}
 AAL acknowledges support from the Isaac Newton Trust, Cambridge, PB
acknowledges support from the EPSRC under grant number GR$/$M88839,

\newpage

\begin{table}[h]
\caption{\label{table1}Best fit coefficients for $v(r;\tilde{\rho})$
as defined in eq 5 for $L=500$ polymers. Coefficients for
$b_3(\tilde{\rho})$ are fixed by the constraint eq 7. The
$\hat{v}(0;\tilde{\rho})$ in eq 7 is approximated by cubic polynomial
$\hat{v}(0;\tilde{\rho})/4\pi = 1.2902 + 0.28132\tilde{\rho} +
0.136761 {\tilde{\rho}}^2 -0.040892 {\tilde{\rho}}^3$ }
\begin{tabular}{lllll}
i &$a_{i0}$ & $a_{i1}$ & $b_{i0}$ & $b_{i1}$\\ \hline 1 & 1.47409 &
-0.07689 & 0.981368 & -0.056808\\ 2 & -0.232096 &0.031321 & 0.42123
&-0.026278 \\ 3 & 0.638974 &0.24193 & - & -\\ \hline
\end{tabular}
\end{table}

\begin{table}[h]
\caption{\label{table2}Best fit coefficients for $v(r;\tilde{\rho})$
as defined in eq 5 for $L=2000$ polymers.  Coefficients for
$b_3(\tilde{\rho})$ are fixed by the constraint eq 7. The
$\hat{v}(0;\tilde{\rho})$ in eq 7 is approximated by cubic polynomial
$\hat{v}(0;\tilde{\rho})/4\pi = 1.245 + 0.3564275 \tilde{\rho}
-0.02443297 {\tilde{\rho}}^2 + 0.0018028 {\tilde{\rho}}^3$ }
\begin{tabular}{lllll}
i &$a_{i0}$ & $a_{i1}$ & $b_{i0}$ & $b_{i1}$\\ \hline 1 &1.41808 &
-0.081969 & 0.979493 & -0.057796 \\ 2 &-0.224377 & 0.030647 & 0.440907
& -0.024327 \\ 3 &0.630219 & 0.211378 & - & -\\ \hline
\end{tabular}
\end{table}

\begin{table}[h]
\caption{\label{table3}Best fit coefficients for
$\phi(z;\tilde{\rho})$ as defined in eq 8 for $L=500$ polymers next to
a wall.  }
\begin{tabular}{llll}
i &$a_{i0}$ & $a_{i1}$ & $a_{i2}$ \\ \hline 0 & 62.7242 & 56.4595&
-29.92825 \\ 1 & -6.40938 & -3.88795 &2.044202 \\ 2 & 2.50812
&5.156190 & -2.13356 \\ 3 & -0.69042 & -1.55191 &0.59725\\ \hline
\end{tabular}

\end{table}

\begin{table}[h]
\caption{\label{tab:rad16}Best fit coefficients for
$\phi(r;\tilde{\rho})$ as defined in eq 10 for $L=500$ polymers around
a sphere of diameter $\sigma = 2R_g$.  }
\begin{tabular}{llll}
i & 0 & 1 \\ \hline $a_{i0}$ & 5.5610 & 1.8477 \\ $a_{i1}$ & -0.8042 &
1.4759 \\ $b_{i0}$ & 0.7751 & 1.2720\\ $b_{i1}$ & -0.1151& 0.1052\\
$c_{i0}$ & 0.4082 & 0.0\\ $c_{i1}$ &0.14104 & 0.0\\ \hline
\end{tabular}
\end{table}

\newpage

\newpage

\section*{List of Figures}

\begin{figure}\epsfig{figure=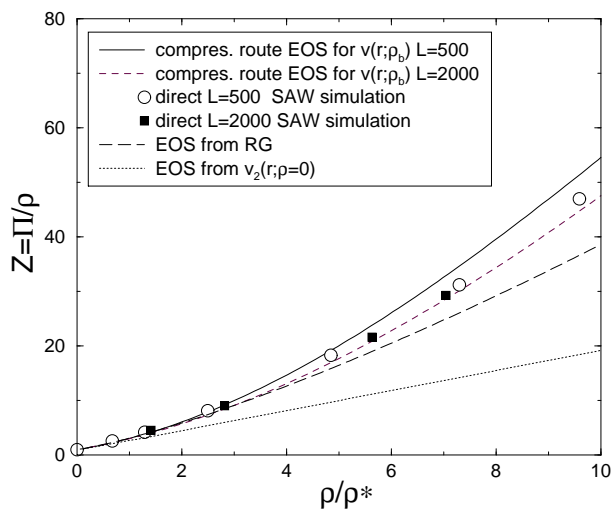,width=8cm} \caption{\label{fig:eos} Linear plot of the EOS\ $Z= \Pi/\rho$ for
$L=500$ and $L=2000$ polymers, here determined from the effective
potentials through the compressibility route and by direct simulations
of the underlying SAW polymer system.  The slight deviations for the
$L=500$ case at the higher densities are most likely due to small
inaccuracies in the inversion procedure used to generate the effective
potentials.  We also show the EOS derived from an
RG\protect\cite{Ohta82} approach.  Using only the $\lim_{\rho
\rightarrow 0}$ form of $v(r;\rho/\rho^*)$ strongly underestimates the
EOS.
}\end{figure}

\end{document}